\documentclass[final,1p,times]{elsarticle}

\usepackage{amssymb}
\usepackage{amsmath}

\usepackage{natbib}
\usepackage{url}
\usepackage{hyperref}
\usepackage{framed}
\usepackage{tcolorbox}
\usepackage{graphicx}
\usepackage{float}
\usepackage{enumitem}
\usepackage{subcaption}

\journal{International Journal of Human-Computer Studies}

\begin{document}

\begin{frontmatter}



\title{Roamify: Designing and Evaluating an LLM Based Google Chrome Extension for Personalised Itinerary Planning}


\author[iiitd]{Vikranth Udandarao\fnref{equal}}
\ead{vikranth22570@iiitd.ac.in}
\author[iiitd]{Noel Abraham Tiju\fnref{equal}}
\ead{noel22338@iiitd.ac.in}
\author[iiitd]{Muthuraj Vairamuthu\fnref{equal}}
\ead{muthuraj22307@iiitd.ac.in}
\author[iiitd]{Harsh Mistry\fnref{equal}}
\ead{harsh22200@iiitd.ac.in}
\author[iiitd,bits]{Dhruv Kumar}
\ead{dhruv.kumar@iiitd.ac.in, dhruv.kumar@pilani.bits-pilani.ac.in}

\affiliation[iiitd]{organization={IIIT Delhi},
            city={New Delhi},
            country={India}}
\affiliation[bits]{organization={BITS Pilani},
            city={Pilani},
            country={India}}

\fntext[equal]{Equal contribution.}

\begin{abstract}
    In this paper, we present Roamify, an Artificial Intelligence powered travel assistant that aims to ease the process of travel planning. We have tested and used multiple Large Language Models like Llama and T5 to generate personalised itineraries per user preferences. Results from user surveys highlight the preference for AI powered mediums over existing methods to help in travel planning across all user age groups. These results firmly validate the potential need of such a travel assistant. We highlight the two primary design considerations for travel assistance: \textbf{D1)} incorporating a web-scraping method to gather up-to-date news articles about  destinations from various blog sources, which significantly improves our itinerary suggestions, and \textbf{D2)} utilising user preferences to create customised travel experiences along with a recommendation system which changes the itinerary according to the user needs. Our findings suggest that Roamify has the potential to improve and simplify how users across multiple age groups plan their travel experiences.
\end{abstract}


\end{frontmatter}



\section{Introduction}
    Traditional planning methods required much information collection and were often restricted to people with access to resources. The general ways of planning a trip included endless web surfing or contacting travel agencies and purchasing one of their travel packages. While these methods produced results, they were not quick or effective enough. All these planning methods have been proven to be tiring and cumbersome. We can employ various AI methodologies to aid us in our travel journey\cite{intro1}.
    
    With the advancements in LLMs and AI, we no longer need to surf the internet continuously to ideate our journey. LLMs have shown their capabilities by trying to understand and replicate human interests through text and other visual representations\cite{intro2}. Over the years, they have shown varied success in different fields, including their ability to provide customised travel-related information.
    
    Roamify utilises LLMs to categorise and provide personalised itinerary generation plans immediately without the users needing to spend their time gathering information.
    
    Exhaustive user surveys and interviews stated that most age groups were eager towards using Roamify to help in their travel planning. Millennials and Generation Z presented positive feedback and looked forward to utilising the platform to aid their travel journey. Surprisingly, Generation X, although accustomed to traditional planning methods, were willing to try the platform for travel planning. Their responses firmly underlined the great deal of potential Roamify has to offer.
    
    To further understand the effectiveness of our work, we have outlined the following practical questions to help guide our research:
    
    \begin{itemize}
        \item \textbf{RQ1 - Current Practices:} How do users plan their travels, and how do they feel about Roamify creating their travel plans employing AI mediums?
        \item \textbf{RQ2 - User Preferences and Personalization:} Would users like to have personalizations in their travel itineraries about their interests, history, adventures, calm sceneries, etc.?
        \item \textbf{RQ3 - Generated Responses:} Are the generated itinerary responses effective and can help their travel journey?
        
    \end{itemize}
    
    Through a set of questions, this paper attempts to illustrate the methodology behind the development of the Roamify application and then find scope within the tourism industry. Additionally, through such in-depth exploration, we wish to present a complete understanding of how mechanisms similar to Roamify can influence and assist in travel planning and aid in capturing human needs and preferences so that user experience can be improved accordingly.

\section{Related Work}
    LLMs have a vast potential to be used for travel purposes, aiding travellers with their planning purposes as they save time and effort. However, a significant issue with using LLMs for travel is needing more tourism knowledge\cite{ref1}. Efforts have been made to fine-tune LLMs to increase their resources for tourism and travel purposes\cite{ref3}. A detailed survey by Shengyu Gu (2024) discusses how these models can be used to explore possibilities such as personalized travel experiences and dynamic itineraries\cite{ref4}. Personalized travel experiences can be achieved by understanding user preferences and sentiments by analyzing textual data from reviews, direct customer interactions, and others. LLMs also possess the ability to generate informative travel guides that keep user preferences in mind, as well as generate compelling descriptions of attractions and sites to visit.
    
    We wanted to achieve the task of personalized itineraries by finding features that capture the user's preferences in the best possible manner. We found diverse natural, amusement, historical, and cultural features. Keeping these genres to a minimum is also important because users must input these details before the itinerary generation begins.

\section{System Design and Architecture}
    
    \begin{figure}[h!]
        \centering
        \includegraphics[width=0.8\textwidth,height=0.4\textheight,keepaspectratio]{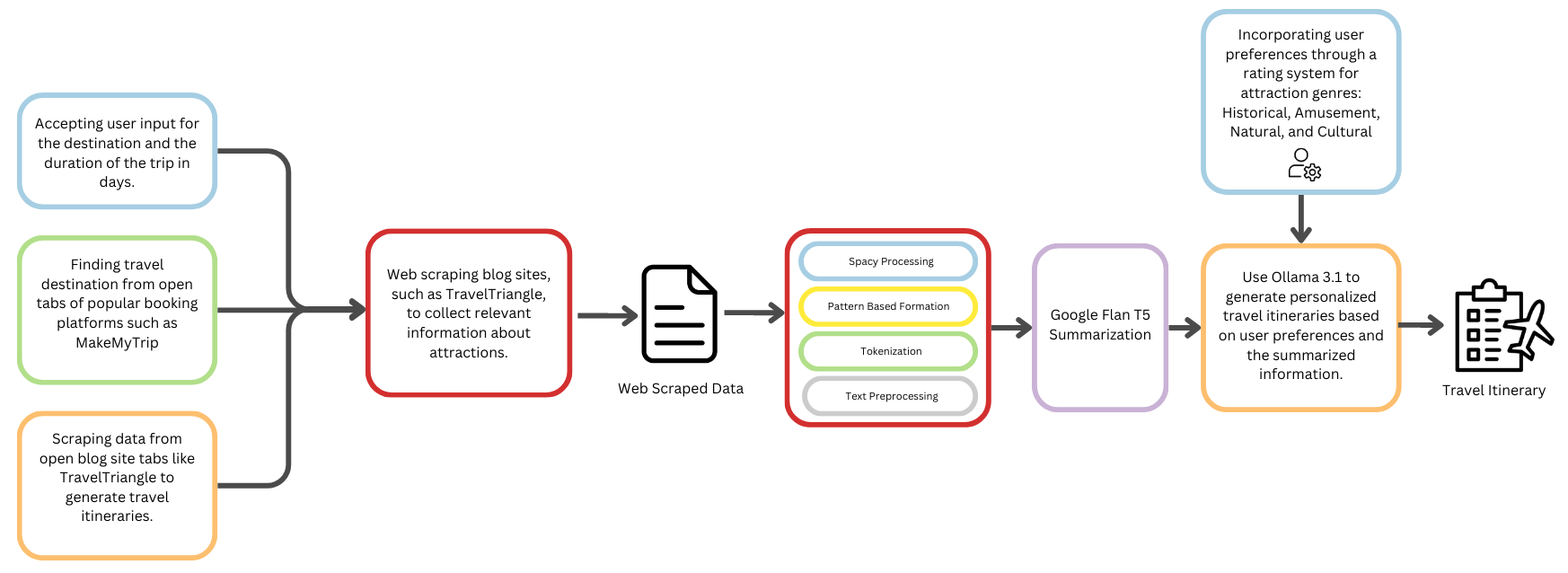}
        \caption{System architecture}
        \label{fig:system-design}
    \end{figure}
    
    Roamify is an LLM-based application designed to generate travel itineraries for users by considering their preferences in different genres and collecting up-to-date data regarding popular travel attractions. As discussed in the previous section, existing LLMs face challenges in planning itineraries as they are trained on relevant data at the time, which may not reflect current travelling trends.
    
    Roamify consists of four primary stages as part of its architecture as described in Figure \ref{fig:system-design}: \textbf{1) Data Collection}, \textbf{2) NLP Processing}, \textbf{3) Summarization}, and \textbf{4) Itinerary Generation}.
    
    \subsection{Data Collection}
        The primary objective of the Data Collection phase is to gather relevant information from travel blogs and related websites. We implemented two methods for determining the user destination.
        
        \begin{itemize}
            \item \textbf{User Input:} The user directly inputs the desired destination and the number of days they plan to spend, thus allowing for a simple and reliable form of user input.
            
            \item \textbf{Web Scraping from Open Tabs:} The extension has
            the ability to identify the destination by analyzing open tabs on the user’s browser. This includes popular travel planning sites and various travel blogs. By scraping these open tabs, the system can infer the user’s intended destination without requiring user input, allowing for a more seamless user experience.
        \end{itemize}
        
        Once the target destination is determined, the system gathers up-to-date information by scraping data from highly-rated travel blog sites. The scraping process is carefully designed to effectively filter advertisements and other miscellaneous information that may not be deemed relevant for our use.
        
        Expanding the scraping corpus comes with two limitations:
        \begin{itemize}
            \item Increased latency
            \item Accumulation of redundant information
        \end{itemize}
    
    \subsection{NLP Processing}
        The second stage in the pipeline involves cleaning and processing the data gathered from the Data collection phase. This stage focuses on extracting only the relevant information about attractions from the scraped text.
        
        Given that different websites follow different structures, we identified a typical pattern across all the sources: attractions are typically numbered in ascending order. Using the insight, we successfully applied the following steps to determine the attraction's name and a short description.
        
        \begin{itemize}
            \item \textbf{Tokenization and Stopword Removal:} First, we tokenize the text and remove stopwords, extracting the sentences.
            
            \item \textbf{Pattern Recognition:} We search for an increasing sequence starting from one. This signifies the start of the list of popular attractions.
            
            \item \textbf{Information Extraction:} The text between consecutive numbers is extracted and considered the description or body for that particular attraction.
            
            \item \textbf{Dictionary Creation:} Finally, we format the extracted information into a dictionary format, where the attractions' names and descriptions are stored as key-value pairs.
        \end{itemize}
        
        This ensures that accurate and organized information is captured on each attraction, laying the basis for our travel itinerary.
    
    \subsection{Summarization}
        This stage of the pipeline deals with condensing the information from the dictionary passed as input from the previous stage. Each key-value pair consists of the attraction name pointing to its details. However, advertisements and other junk text may still be present in these descriptions if they were not filtered out successfully by the previous stages. Clearing this persistent junk requires a more advanced tool such as Google Flan T5 or Llama-8b-Instruct models. Before fine-tuning the model, the task of summarization was performed adequately; however, to reduce the overall latency of the process, we required only relevant details to be passed on to the last stage.
        
        To achieve this, we fine-tuned both models using a custom dataset we prepared, which consisted of two components:
        
        \begin{itemize}
          \item \textbf{Context:} The attraction details extracted from the dictionary.
          \item \textbf{Summary:} The concise and relevant summary of the attraction details.
        \end{itemize}
        
        The following example demonstrates the context obtained after scraping, the summary used in the fine-tuning datasets, and the outputs generated by T5 and LLaMA, respectively:
    
    \begin{tcolorbox}[boxrule=1pt, left=15pt, right=15pt, top=0pt, bottom=0pt]
    
    \textbf{Context:}
    
    Cubbon Park, Sarangib for Pixabay. Situated over a sprawling 300 acres of land, the park was constructed by Richard Sankey. This massive green park, along with lawns, deserves a special mention. Offering statues of famous personalities, the park is one among the popular places to visit in Bangalore with friends. Location: Kasturba Road, Behind High Court of Karnataka, Ambedkar Veedhi, Sampangi Rama Nagara, Bangalore. Timings: Open on all days. Entry Fee: No entry fee. Suggested Read: Resorts Near Bangalore.
    \end{tcolorbox}
    \begin{tcolorbox}[boxrule=1pt, left=15pt, right=15pt, top=0pt, bottom=0pt]
    
    \textbf{Summary (Dataset):}
    
    Cubbon Park, spanning 300 acres and constructed by Richard Sankey, is a massive green space in Bangalore featuring lawns and statues of famous personalities. Located on Kasturba Road, it is a popular spot for outings with friends. The park is open daily with no entry fee.
    \end{tcolorbox}
    \begin{tcolorbox}[boxrule=1pt, left=15pt, right=15pt, top=0pt, bottom=0pt]
    
    \textbf{T5 Output:}
    
    Cubbon Park, a sprawling 300 acres of land, is a popular place to visit in Bangalore with friends. Offering statues of famous personalities, it is a popular place to visit with friends.
    \end{tcolorbox}
    \begin{tcolorbox}[boxrule=1pt, left=15pt, right=15pt, top=0pt, bottom=0pt]
    
    \textbf{LLaMA Output:}
    
    The park was constructed by Richard Sankey. This massive green park, along with lawns, deserves a special mention. Offering statues of famous personalities, the park is one among the popular places to visit in Bangalore with friends. Location: Kasturba Road, Behind High Court of Karnataka.
    \end{tcolorbox}
    
    From the above descriptions, we can see that T5 is more effective at capturing the essential information; hence, it successfully condenses the description while clearing out any junk text. On the other hand, Llama was more successful at elaborating on the description. However, this came with a much higher processing time than T5. Since processing time is an essential factor for our application, we integrated the T5 transformer into the pipeline as our mode for summarization.
    
    \subsection{Itinerary Generation}
        The final stage of the pipeline involves using Llama-3 or ChatGPT-4 to plan the itinerary based on the summarized and concise attractions given to it by the T5 transformer. The model then plans an itinerary using the attractions based on the number of days inputted by the user.
        
        \begin{tcolorbox}[boxrule=1pt, left=15pt, right=15pt, top=15pt, bottom=15pt]
        
        \textbf{Prompt Design:} \\
        Generate a detailed itinerary for me for a \underline{\textit{days}} day trip to  \underline{\textit{destination}} and here are the suggested places I would like to cover:
        
          \begin{itemize}
              \item \textbf{1. Attraction Name:}
              \begin{itemize}
                  \item \textit{Description:} Attractions Details
              \end{itemize}
              \item \textbf{2. Attraction Name:}
              \begin{itemize}
                  \item \textit{Description:} Attractions Details
              \end{itemize}
              \item \textbf{N. Attraction Name:}
              \begin{itemize}
                  \item \textit{Description:} Attractions Details
              \end{itemize}
          \end{itemize}
        \end{tcolorbox}
        
        The prompt shown above is used to generate a general itinerary using Ollama.
        
        However, these itineraries still needed to achieve the purpose of personalized recommendations based on users' preferences. On interviewing users, we identified some key features or distinct attraction genres that are important to them:
        
        \begin{itemize}[noitemsep, topsep=5pt, parsep=0pt, partopsep=0pt]
            \item \textbf{Historical}
            \item \textbf{Amusement}
            \item \textbf{Natural}
            \item \textbf{Cultural}
        \end{itemize}
        
        For each genre, the user is shown a slider from 1 to 5 to rate it. These ratings are stored and added to the prompt so the itinerary can be planned accordingly.
    
    
    \begin{figure}[H]
        \centering
        \includegraphics[width=0.95\columnwidth,height=0.6\textheight]{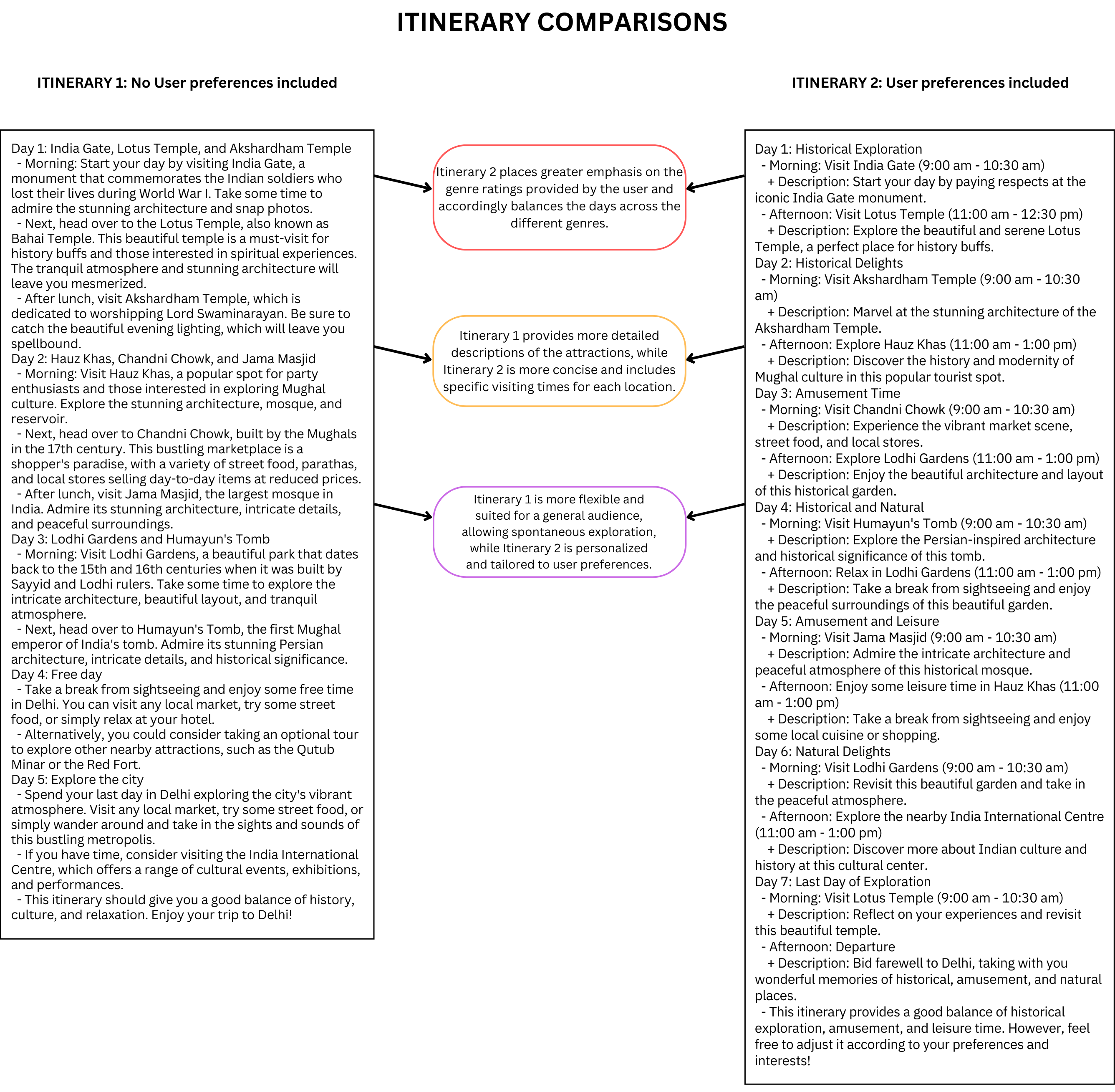}
        \caption{Comparison of Itineraries With and Without User Preferences Included}
        \label{fig:itinerary_comparison}
    \end{figure}
    
    \subsection{Comparison of the Itineraries}
        
        In Figure \ref{fig:itinerary_comparison}, a detailed comparison between two itineraries is presented: one that does not take user preferences into account (Itinerary 1) and another that does (Itinerary 2). There are three salient differences that can be observed between the two itineraries:
    
        \begin{itemize}[noitemsep, topsep=5pt, parsep=0pt, partopsep=0pt]
        \item \textbf{Genre Emphasis:} Itinerary 2 places greater emphasis on the genre ratings provided by the user and accordingly balances the days across the different genres.
        
        \item \textbf{Detail Level:} Itinerary 1 provides more detailed descriptions of the attractions, while Itinerary 2 is more concise and includes specific visiting times for each location.
        
        \item \textbf{Flexibility vs. Personalization:} Itinerary 1 is more flexible and suited for a general audience, allowing spontaneous exploration, while Itinerary 2 is personalized and tailored to user preferences.
        \end{itemize}
        
        Keeping these differences in mind, we sought to find which itinerary is more beneficial for our use case. On asking the interview participants, 75\% chose the itinerary that considered user preferences and appreciated the level of personalization it offered. Some participants were even told to go beyond the selected four features: historical, natural, amusement, and cultural, and suggest additional valuable features.

\section{User Survey}
    
\begin{figure}[h!]
    \centering
    \begin{subfigure}[b]{0.22\textwidth} 
        \centering
        \includegraphics[height=2cm]{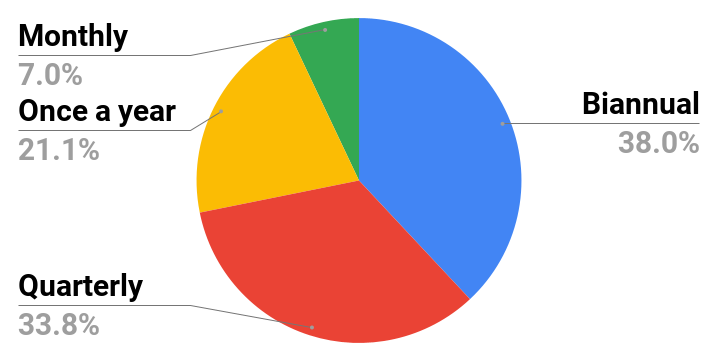} 
        \caption{Travel frequency}
        \label{fig:travel-freq}
    \end{subfigure}
    \hfill
    \begin{subfigure}[b]{0.22\textwidth} 
        \centering
        \includegraphics[height=2cm]{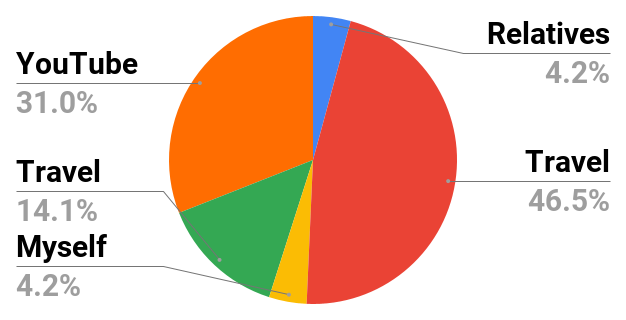} 
        \caption{Itinerary Decision}
        \label{fig:itinerary-dec}
    \end{subfigure}
    \hfill
    \begin{subfigure}[b]{0.22\textwidth} 
        \centering
        \includegraphics[height=2cm]{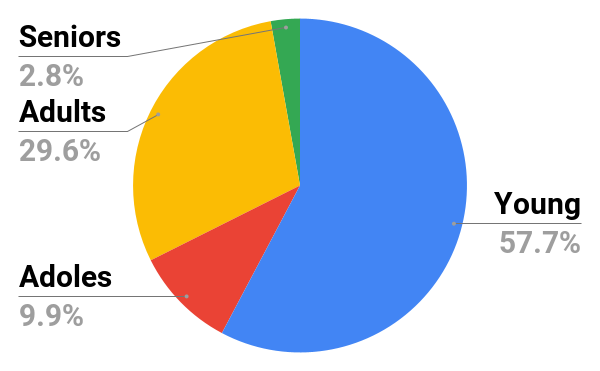} 
        \caption{AI Planner Preferences}
        \label{fig:preference-planner}
    \end{subfigure}
    \hfill
    \begin{subfigure}[b]{0.22\textwidth} 
        \centering
        \includegraphics[height=2cm]{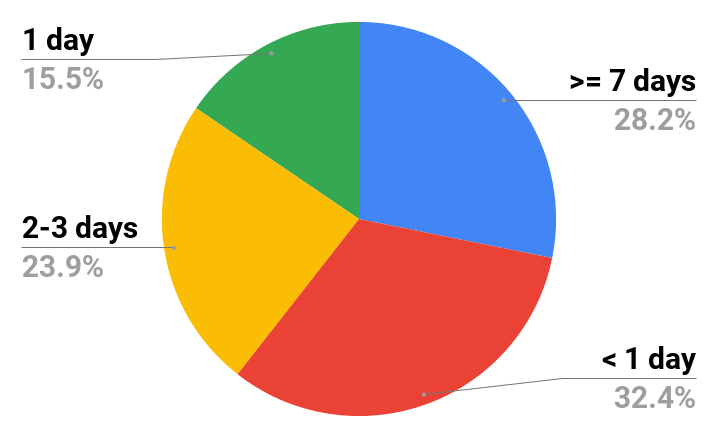} 
        \caption{Itinerary Build Time}
        \label{fig:itinerary-time}
    \end{subfigure}
    \caption{Visualizations of user survey responses}
    \label{fig:1x4-grid-wide}
\end{figure}
    
    To understand user preferences and travel behaviours, we surveyed 200 individuals of various demographic groups and visualized the results in a series of pie charts through means like google forms and questionnaires. The survey included multiple-choice and open-ended questions to capture insights into travel habits, planning challenges, and expectations from a travel planning tool.
    Below are the key findings:
    
    \begin{itemize}
        \item \textbf{Types of Travel}: Casual travel was most common(38\%), indicating a preference for non-work-related trips. Entertainment-related travel also made up a significant portion (29.6\%), followed by work-related trips (11.3\%). Family trips were less common, constituting only 2.8\%, while minimal travel was motivated by meetups (8.5\%) or taking a break (1.4\%).
        
        \item \textbf{Travel Frequency}: The most common travel frequency reported by participants was once every six months as seen in \ref{fig:travel-freq}.
        
        \item \textbf{Satisfaction with Travel Agencies}: Respondents were largely neutral about their experience with travel agencies, with 43.8\% selecting neutral. Those satisfied made up 29.7\%, while a small portion expressed being very satisfied (6.3\%). Dissatisfaction was also present, with 10.9\% feeling unsatisfied and 9.4\% being very unsatisfied.
        
        \item \textbf{Itinerary Decision}: Online travel websites were the most popular source for itinerary planning as seen in \ref{fig:itinerary-dec}.
        
        \item \textbf{Help from Friends/Family for Itinerary}: A striking 88.7\% of respondents indicated that they seek help from friends or family for planning their itineraries, while only 11.3\% prefer doing it alone.
        
        \item \textbf{Demographic Distribution}: Young adults (ages 19-25) made up the majority of participants as seen in \ref{fig:preference-planner}.
        
        \item \textbf{Customization Preferences}: A large majority of respondents (88.7\%) preferred to customize their own itineraries rather than rely on preselected ones (7\%), with a small group opting for a mix of both (4.2\%).
        
        \item \textbf{Custom Itinerary Build Time}: A slightly higher percentage of participants took less than 1 day to build an itinerary as seen in \ref{fig:itinerary-time}.
    \end{itemize}
    
    This detailed user survey analysis highlights the travel preferences and behaviors of the respondents, offering insights that can be used to tailor future travel planning services.

\section{Interview Analysis and Testimonials}
    
    User surveys, interviews and recorded testimonials serve as the main bridge connecting an upcoming technology to its user space. To ensure Roamify serves the user's needs and ideally serves as a proper tool for navigation, around 10 interviews were conducted with travel influencers across the globe to understand the true potential of the product. Many travel influencers, bloggers, and frequent travellers were reached out; which can be accessed \href{https://drive.google.com/drive/folders/1mKPTXZ7n7ZFmMEK6QBOKU0op3mf8hCta?usp=sharing}{\textbf{here}}. Profiles were scouted by web surfing techniques, including approaching via social media profiles, blog channels and contacting associated managers.
    
    Our outreach included messages and formal emails, to set up quick interviews to ask questions related to the travel domain. The questionnaire focused on how frequent travellers plan their trips and whether they approach any concerned travel agency to aid in their travelling journey. The results were enthralling; most travel influencers preferred finding out more about locations to travel to and plan their trips on their own, mainly by web surfing. Many preferred to curate information by themselves and opted for impromptu trips and sudden planning.
    
    To deepen our understanding, questions were asked regarding the process behind planning trips. Their planning focused on finding exotic locations that other travellers rarely explored. Furthermore, travellers were interested in discovering new places and also were concerned with finding the logistics behind each trip.
    
    In light of these responses above, further questions were asked about the potent travelling hurdles travellers had encountered. Many travel influencers felt that the current online data is outdated and does not reflect the authentic travel experiences they seek. Furthermore, many influencers exclaimed they would like to find more accessible and affordable stays, along with itineraries, instantly.
    
   After analysing user needs, we proposed an AI Travel assistance that will firmly aid travellers in travel planning processes. Furthermore, we presented the concept of personalisation, where users can plan and personalise their journey according to their interests. For instance, history enthusiasts can personalize to visit more historical locations.
    
    Responses in the light of this matter were alarmingly positive. Most of the travellers had been utilising AI generated resources to help in their day-to-day tasks; when proposed with the same AI-based medium, which would help them in their travel journey and save them hours and hours of information scouting, they were affirmative towards the notion. The personalisation feature further enhanced the appeal for the AI-based travel medium.
    
    Furthermore, Roamify integrates Llama 3, ensuring real-time updates and precise information. Challenges pertaining to accommodation were addressed as one of the core features in Roamify's development pipeline ensuring the AI-based medium is user friendly and accessible.
    
    Henceforth, we presented our prototype and received valuable feedback. The strengths included the user-friendly GUI of the platform, and the recommendation system that carefully personalizes and crafts itineraries to align with user's needs. Suggestions for improvement included adding more logistical details and stay-related information to help create an enriching experience. From the diverse survey conducted with travel influencers of various demographics, it was observed that despite the age gap, participants were enthusiastic about using an AI mechanism to meet their travel needs and create a better experience with their loved ones.

\section{Conclusion}
    
    The goal of this research was to identify the best combination of LLMs for generating and fine-tuning itineraries. Initially, we trained question-answering models to extract details from attraction descriptions, but they underperformed. We then focused on text-to-text generation models, specifically LLaMA-3 and T5, for summarization.
    
    Our experiments showed that while T5 is faster at summarization, it is sometimes less detailed than LLaMA-3. Given our priority on speed, we selected T5 for summarization, pairing it with LLaMA for its detailed output. Ollama, based on LLaMA 3.1, was used for efficient itinerary generation.
    
    We also investigated user preferences in itinerary planning, integrating them into Ollama's prompt for more personalized and satisfactory travel plans.
    
    In conclusion, the combination of T5 and LLaMA provided a balanced approach, ensuring both efficiency and comprehensiveness, while incorporating user preferences significantly enhanced personalization in travel itineraries.
    
    \subsection{Future Work}
        
        This study demonstrates the potential of advanced NLP techniques in transforming travel planning. The successful implementation in Roamify shows these models can generate accurate and personalized itineraries. Future work will aim to optimize model performance and efficiency, focusing on:
        \begin{itemize}
            \item \textbf{Data Collection and Processing:} Explore reliable data collection strategies beyond web scraping for more reliability and also utilize advanced NLP models for better processing.
            \item \textbf{User-Centric Enhancements:} Integrate user feedback to refine itinerary recommendations and offer options to save itineraries directly to calendars for convenience.
            \item \textbf{Application Expansion:} Extend model capabilities to real-time travel assistance, dynamic itinerary adjustments, and personalized services tailored to traveler profiles based on cost, comfort, and time priorities.
        \end{itemize}

\section*{Acknowledgment}

    We extend our sincere gratitude to Dr. Anmol Srivastava (Human-Centered Design Department, IIIT-Delhi), Vishaal Udandarao (University of Tübingen), and Arpit Bhatia (University of Copenhagen) for their valuable feedback and insightful suggestions. Their expertise and thoughtful reviews have been instrumental in refining the ideas and improving the overall quality of this work.

\newpage

\bibliographystyle{plain}

\newpage

\appendix

\section{User Survey}
    To understand the travel planning behaviours and preferences of potential users for Roamify, we conducted a comprehensive survey. Below are the survey questions and the options provided for multiple-choice questions.

    \begin{enumerate}
        \item \textbf{What is your name?}

        \item \textbf{Which age group do you belong to?}
        \begin{itemize}
            \item Middle Childhood (6-11 years)
            \item Adolescents (12-18 years)
            \item Young Adults (19-25 years)
            \item Adults (26-64 years)
            \item Seniors (65 years+)
        \end{itemize}

        \item \textbf{How frequently do you travel?}
        \begin{itemize}
            \item Once a month
            \item Once every 3 months
            \item Once every 6 months
            \item Once a year
            \item Other
        \end{itemize}

        \item \textbf{What type of travel you usually undergo?}
        \begin{itemize}
            \item Work
            \item Entertainment
            \item Casual
            \item Meetup
            \item Other
        \end{itemize}

        \item \textbf{How do you decide your itinerary?}
        \begin{itemize}
            \item Travel Agents
            \item YouTube videos
            \item Online Travel websites
            \item Other
        \end{itemize}

        \item \textbf{How do you decide your itinerary?}
        \begin{itemize}
            \item Travel Agents
            \item YouTube videos
            \item Online Travel websites
            \item Other
        \end{itemize}

        \item \textbf{Do you like to travel based on your own customized itinerary or preselected recommended tourist packages?}
        \begin{itemize}
            \item Own Customized
            \item Preselected Recommended Tourist Packages
            \item Other
        \end{itemize}

        \item \textbf{If you have used a travel agency how much are you satisfied with their itinerary?}
        \begin{itemize}
            \item 1
            \item 2
            \item 3
            \item 4
            \item 5
        \end{itemize}

        \item \textbf{If you make custom itinerary how long do you take to build one?}
        \begin{itemize}
            \item Open-ended response
        \end{itemize}

        \item \textbf{Do you ask your friends and family for help who are currently living or have visited that place?}
        \begin{itemize}
            \item Yes
            \item No
        \end{itemize}
    \end{enumerate}

    \section{Interview Questions}
    In addition to the survey, we conducted in-depth interviews with selected participants to gain deeper insights into their travel planning experiences. Below are the interview questions:

    \begin{enumerate}
        \item How frequently do you travel?
        \item Can you describe your typical process for planning a trip?
        \item What challenges do you face when planning a trip online?
        \item Have you ever used an offline travel agency? How was your experience compared to planning trips online?
        \item What features would you find most helpful in a travel planning tool?
        \item How important is the personalization of travel recommendations to you?
        \item Can you provide an example of a travel planning experience that was particularly stressful or frustrating?
        \item How do you think AI can improve the travel planning experience?
        \item What do you expect from a travel planning app or tool in terms of user experience?
        \item Have you ever used an AI-driven travel planning website where you can ask for personalized travel suggestions? If so, how was your experience?
    \end{enumerate}

\end{document}